\documentclass[%
 reprint,
 amsmath,amssymb,
 aps,
]{revtex4-2}
\usepackage{graphicx}
\usepackage{dcolumn}
\usepackage{bm}
\usepackage{color}
\usepackage{ulem}

\begin{document}

\preprint{APS/123-QED}
\title{Folding dynamics and its intermittency in turbulence}


\author{Yinghe Qi$^1$}

\author{Charles Meneveau$^1$}

\author{Greg Voth$^2$}


\author{Rui Ni$^1$}
 \email{rui.ni@jhu.edu}

\affiliation{$^1$ Department of Mechanical Engineering, The Johns Hopkins University, Baltimore, MD 21218, USA\\
$^2$ Department of Physics, Wesleyan University, Middletown, CT 06459, USA}

\date{\today}

\begin{abstract}
Fluid elements deform in turbulence by stretching and folding. In this work, by projecting the material deformation tensor onto the largest stretching direction, the dynamics of folding is depicted through the evolution of the material curvature. Results from direct numerical simulation (DNS) show that the curvature growth exhibits two regimes, first a linear stage dominated by folding fluid elements through a persistent velocity Hessian which then transitions to an exponential growth driven by the stretching of already strongly bent fluid elements. This transition leads to strong curvature intermittency at later stages, which can be explained by a proposed curvature-evolution model. The link between velocity Hessian to folding provides a new way to understand the crucial steps in energy cascade and mixing in turbulence beyond the classical linear description.




\end{abstract}

\maketitle

The deformation of fluid elements, as already described by Reynolds in 1894 \cite{reynolds1894study}, is a process that involves stretching and folding. Stretching elongates fluid elements exponentially \cite{haller2015lagrangian} along one (or two) direction(s) and compresses them in the other directions, while folding brings fluid particles closer, which increases the local curvature and also reduces length scales.  Given its connection to flow structures and their dynamics, deformation is therefore essential to many fundamental problems in turbulence including mixing \cite{shraiman_scalar_2000,villermaux_mixing_2019}, energy cascade \cite{meneveau1991multifractal}, and vortex dynamics \cite{guala2005evolution}, as well as in turbulent multiphase flows with non-spherical \cite{ni2014alignment,zhao2019passive} and deformable particles \cite{magnaudet2000motion,qi2022fragmentation}.


The linear component of deformation has been studied extensively in turbulence \cite{cocke1969turbulent,girimaji_material-element_1990,dresselhaus1992kinematics,villermaux1994line,voth2002experimental,haller2015lagrangian}, and the dynamic equation for linear deformation links the geometries of flow structures to the velocity gradient and Cauchy-Green strain tensors. This linkage paves the foundation to finite-time Lyapunov exponent and the Lagrangian coherent structures \cite{haller2015lagrangian}, which have impacted studies of the transport and mixing of passive scalars in the atmosphere \cite{sherwood_spread_2014,garratt1994atmospheric}, ocean \cite{vic_deep-ocean_2019}, and solar interior \cite{schumacher2020colloquium}

The natural question arises as to how such a framework can be extended to the folding dynamics and what is the right dynamical system approach for describing folding. Given the nonlinear nature of problem, several different methods have been proposed, such as taking the total deviation from the linear part, \cite{kelley2011separating} or calculating the curvature of fluid elements \cite{girimaji_material-element_1990,girimaji1991asymptotic,ma2016stretching,bentkamp2022statistical}. However, the connection from these statistics to the underlying fluid dynamics in the Eulerian and Lagrangian frameworks have not been clearly illustrated.

\color{black}





\begin{figure}
\centering
\includegraphics[width=0.9\linewidth]{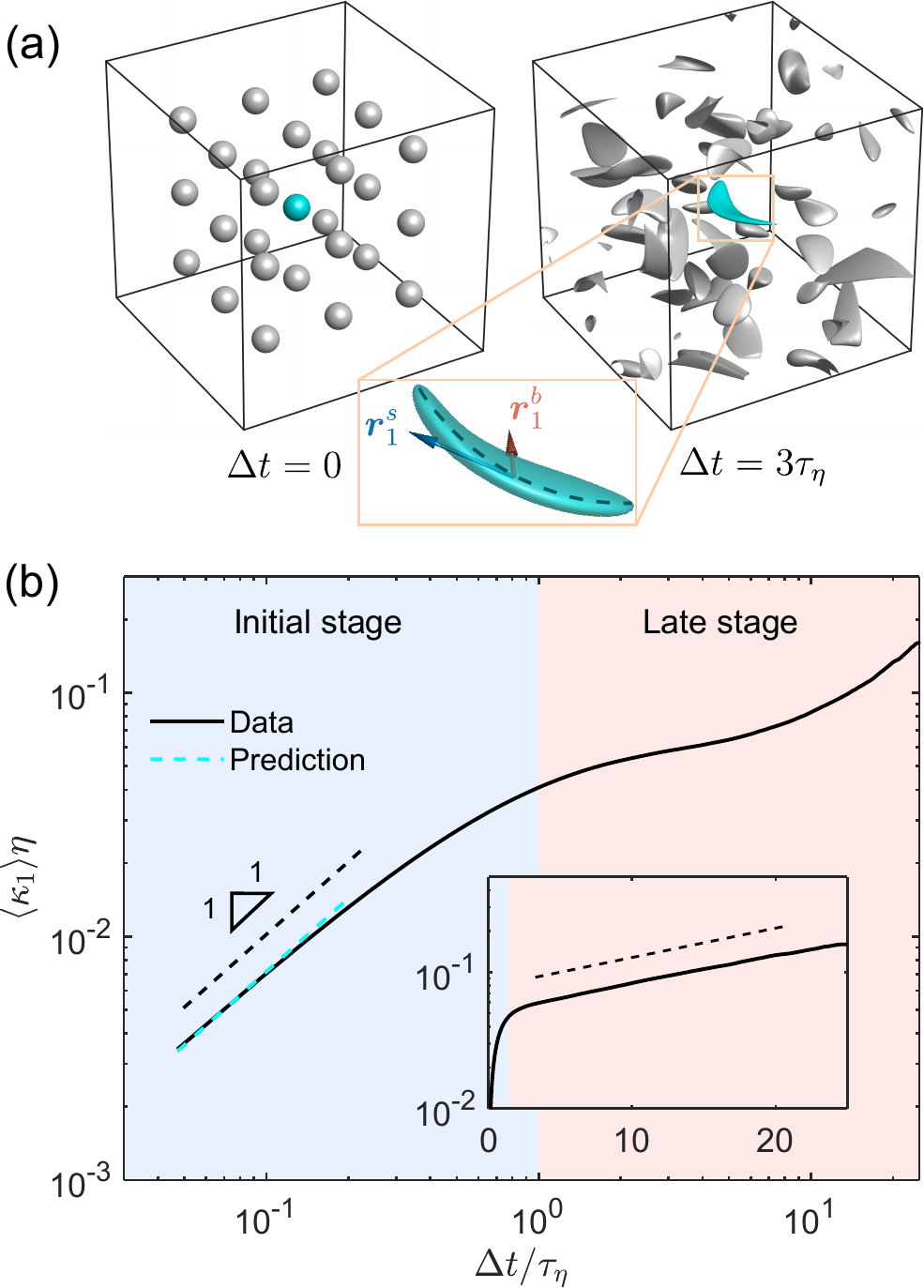}
\caption{(a) The deformation of infinitesimal spherical fluid elements after $\Delta t=3\tau_\eta$ (size not to scale). (b) The time evolution of the mean curvature $\langle\kappa_1\rangle$ (black solid curve); The black dashed line represents a linear relationship, and the cyan dashed line represents the prediction based on an Eulerian quantity $\langle|\boldsymbol{\hat{e}}_1\cdot\boldsymbol{H}\cdot\boldsymbol{\hat{e}}_1|\rangle$ with $\boldsymbol{\hat{e}}_1$ being the eigenvector corresponding to the maximum eigenvalue of the rate-of-strain tensor. Inset: the same figure for $\langle\kappa_1\rangle$ (black solid line) with a linear scale in time. The black dashed line represents an exponential growth over time.\label{fig_element_deformation}}
\end{figure}

To build a framework that makes that connection, we consider the folding of infinitesimal fluid elements. Fig. \ref{fig_element_deformation}(a) shows a number of infinitesimal spherical fluid elements being deformed after a time $3\tau_\eta$ ($\tau_\eta$ is the Kolmogorov time scale) in 3D homogeneous and isotropic turbulence \cite{li2008public,perlman2007data} (details of the direct numerical simulation (DNS) of the turbulence can be found in Supplemental Material). It is clear that the deformed fluid elements show complex geometry involving both stretching and folding. To mathematically describe this high-order deformation, we consider each point $\boldsymbol{X}$ at $t_0$ within an infinitesimal fluid element mapped to another point $\boldsymbol{x}$ within the deformed element after a finite time $\Delta t$, where $\boldsymbol{x}$ and $\boldsymbol{X}$ are the relative positions with respect to the center of the fluid elements. The non-linear mapping function between $\boldsymbol{X}$ and $\boldsymbol{x}$ with the leading orders follows
\begin{equation}\label{eqn_mapping}
\boldsymbol{x}=\boldsymbol{F}(t_0+\Delta t)\cdot\boldsymbol{X}+\boldsymbol{X}\cdot\boldsymbol{G}(t_0+\Delta t)\cdot\boldsymbol{X},
\end{equation}
where $F_{ij}=\partial x_i/\partial X_j$ is the deformation gradient tensor and $G_{ijk}=\partial^2 x_i/\partial X_j\partial X_k$ is the deformation Hessian tensor. The tensors $F_{ij}$ and $G_{ijk}$ can be then determined by integrating  $dF_{ij}(t)/dt=A_{im}F_{mj}(t)$ and  $dG_{ijk}(t)/dt=A_{im}G_{mjk}(t)+H_{imn}F_{mj}(t)F_{nk}(t)/2$ along the trajectories of fluid elements, with $A_{ij}=\partial u_i/\partial x_j$ and $H_{ijk}=\partial^2 u_i/\partial x_j \partial x_k$ being the velocity gradient and velocity Hessian tensors, respectively. Details of these equations can be found in Supplemental Material. 

To further simplify Eq. (\ref{eqn_mapping}), we consider the deformation of an arbitrary straight material line passing through the center of a fluid element, represented by a set of positions $\boldsymbol{X}$ represented parametrically according to $\boldsymbol{X}(\lambda)=\boldsymbol{\hat{e}}\lambda$.  $\boldsymbol{\hat{e}}$ is a selected unit vector and the parameter $\lambda\rightarrow0$ indicates the distance from the center of the fluid element. Substituting $\boldsymbol{X}(\lambda)=\boldsymbol{\hat{e}}\lambda$ into Eq. (\ref{eqn_mapping}) yields the expression for the deformed material line at $t_0+\Delta t$, 
\begin{equation}\label{eqn_mapping_expand}
    \boldsymbol{x}(\lambda)=\boldsymbol{F}\cdot \boldsymbol{\hat{e}} \lambda+\boldsymbol{\hat{e}}\cdot\boldsymbol{G}\cdot\boldsymbol{\hat{e}}\lambda^2= \boldsymbol{r}^s \lambda+\boldsymbol{r}^b \lambda^2,
\end{equation}
where $\boldsymbol{r}^s=\boldsymbol{F}\cdot \boldsymbol{\hat{e}}$ and  $\boldsymbol{r}^b=\boldsymbol{\hat{e}}\cdot\boldsymbol{G}\cdot\boldsymbol{\hat{e}}$ are defined as the stretching vector and the bending vector, respectively. 



A highly relevant material line is the one that gets stretched the most, written as $\boldsymbol{X}(\lambda)=\boldsymbol{\hat{e}}_{R1}\lambda$. Here, $\boldsymbol{\hat{e}}_{R1}$ is the unit eigenvector associated with the greatest eigenvalue of right Cauchy-Green strain tensor $\boldsymbol{C}^R=\boldsymbol{F}^T\boldsymbol{F}$. This special material line, as the "skeleton" of the fluid element, can be used to reflect the overall geometry of the fluid element. Substituting $\boldsymbol{\hat{e}}=\boldsymbol{\hat{e}}_{R1}$ in Eq. (\ref{eqn_mapping_expand}) results in the quadratic equation $\boldsymbol{x}(\lambda)= \boldsymbol{r}^s_1 \lambda+\boldsymbol{r}^b_1 \lambda^2$ where $\boldsymbol{r}^s_1=\boldsymbol{F}\cdot\boldsymbol{\hat{e}}_{R1}$ and $\boldsymbol{r}^b_1=\boldsymbol{\hat{e}}_{R1}\cdot\boldsymbol{G}\cdot\boldsymbol{\hat{e}}_{R1}$. An example of this material line is shown as the inset of Fig.  \ref{fig_element_deformation}(a) (black dashed line). Given this quadratic equation, the curvature of the material line $\kappa_1$ can be found using $\kappa_1=2r^b_{1\perp}/(r_1^s)^2$, where $\boldsymbol{r}^b_{1\perp}$ represents the component of $\boldsymbol{r}^b_1$ that is perpendicular to $\boldsymbol{r}^s_1$. Although $\kappa_1$ is not sufficient to describe the complete deformation, it does reflect the overall folding of the fluid element.

The curvature $\kappa_1$ can therefore be obtained by computing $\boldsymbol{F}$ and $\boldsymbol{G}$ and their associated $\boldsymbol{r}^b_1$ and $\boldsymbol{r}^s_1$ along with each fluid trajectory. Fig. \ref{fig_element_deformation}(b) shows the time evolution of the mean curvature $\langle\kappa_1\rangle$, averaged over $10^5$ fluid elements, as a function of the integration time $\Delta t$ using the DNS data. It is evident that, for the available simulation duration, the mean curvature of the fluid elements grows continuously, but the growth rate changes appreciably between two regimes. In early times, $\langle\kappa_1\rangle$ increases linearly. The linear regime lasts until about the Kolmogorov timescale $\tau_\eta$ when the length scale $1/\langle\kappa_1\rangle$ is around 25$\eta$ ($\eta$ is the Kolmogorov length scale), and the growth of $\langle\kappa_1\rangle$ slows down, marking the transition of the curvature dynamics. Soon after $\tau_\eta$, the growth of  $\langle\kappa_1\rangle$ accelerates again, and this late stage behavior is better fitted with an exponential function, which is illustrated in a semi-logarithmic plot in the inset of Fig. \ref{fig_element_deformation}(b). 


\begin{figure}
    \centering
    \includegraphics[width=0.9\linewidth]{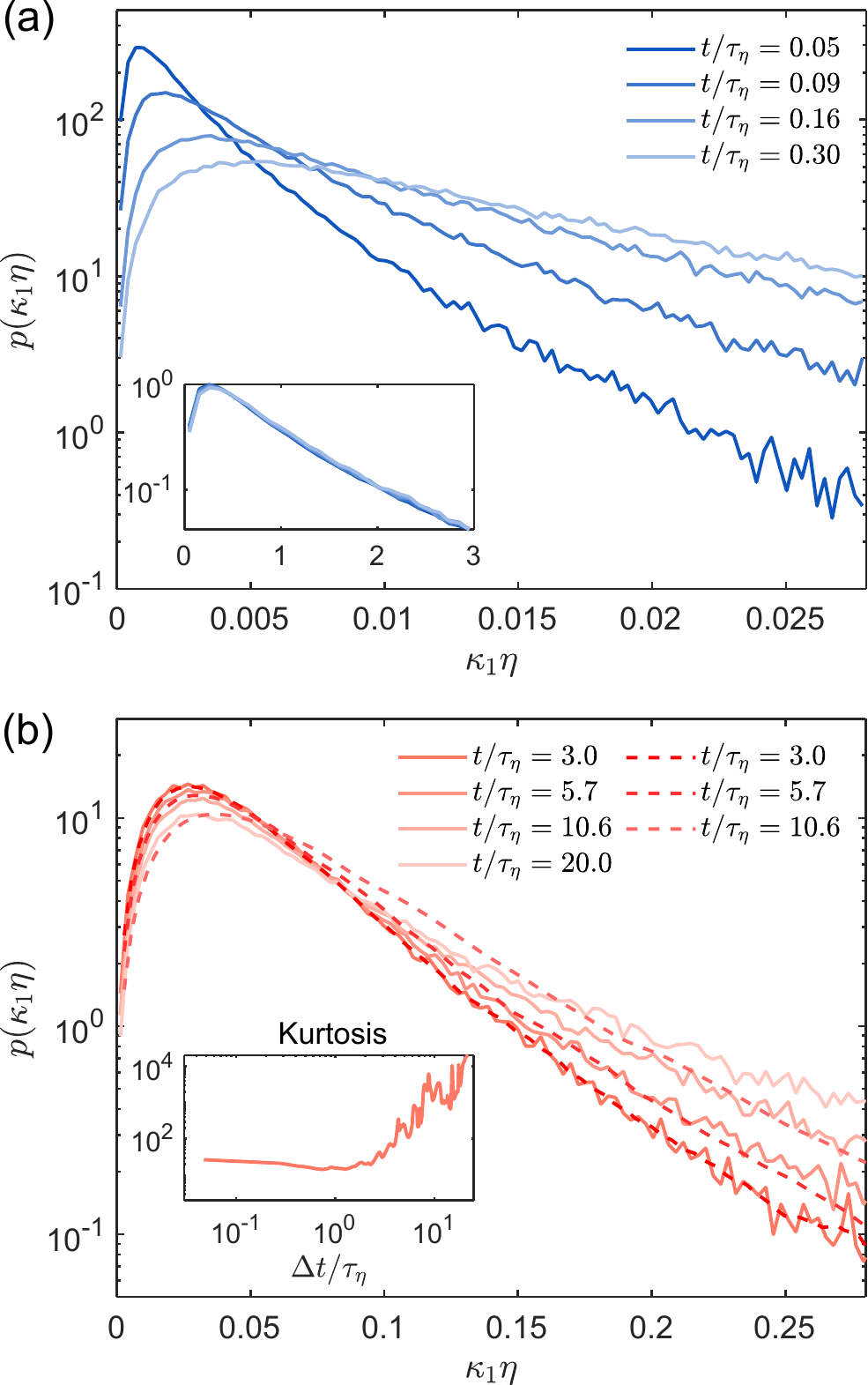}
    \caption{(a) The PDFs of the curvature $p(\kappa_1)$ at different time instants in the early stage. Inset of (a): the same PDFs but for the normalized curvature $p(\kappa_1/\langle\kappa_1\rangle)$. (b) The PDFs of the curvature $p(\kappa_1)$ at different time instants in the late stage with the solid curves representing the data and the dashed curves representing the prediction by the model (Eq. (\ref{eqn_pdf_evolution})). Inset of (b): the time evolution of the kurtosis of $\kappa_1$. }
    \label{fig_curv_pdf}
\end{figure}

The transition from the linear to the exponential growth of $\langle \kappa_1 \rangle$ indicates different mechanisms at play, which can be better understood using local curvature. Here, the probability density function (PDF) of $\kappa_1$, i.e. $p(\kappa_1)$, at different times  are shown in Fig. \ref{fig_curv_pdf} for the early (a) and late (b) stages. In the early stage, the curvature grows systematically, but follows a self-similar behavior as indicated by the collapsed PDFs of the normalized curvature $p(\kappa_1/\langle \kappa_1 \rangle)$ in the inset of Fig. \ref{fig_curv_pdf}(a). In the late stage, the tail of the PDF still rises over time, whereas the peak location remains constant. This distinct behavior suggests that the curvature distribution becomes more intermittent over time, which is confirmed by the growing kurtosis as shown in Fig. \ref{fig_curv_pdf}(b) inset. This result highlights the growing inhomogeneity of local mixing as locations with extreme curvature should reach a well-mixed stage much sooner than what is implied by the mean.




To model the multi-stage growth behavior of curvature, we consider an arbitrary deforming infinitesimal material line as in Eq. (\ref{eqn_mapping_expand}). The equation for this material line can therefore be decomposed along two directions, $\boldsymbol{\hat{e}}_\parallel=\boldsymbol{r}^s/r^s$ and $\boldsymbol{\hat{e}}_\perp=\boldsymbol{r}^b_\perp/r^b_\perp$ , following:
\begin{equation}\label{eqn_geometry}
    \boldsymbol{x}(\lambda)=\left(r^s\lambda+r^b_\parallel\lambda^2\right)\boldsymbol{\hat{e}}_\parallel+r^b_\perp\lambda^2\boldsymbol{\hat{e}}_\perp,
\end{equation}
where $\boldsymbol{r}^b_\parallel=(\boldsymbol{r}^b\cdot\boldsymbol{\hat{e}}_\parallel)\boldsymbol{\hat{e}}_\parallel$ and $\boldsymbol{r}^b_\perp=\boldsymbol{r}^b-\boldsymbol{r}^b_\parallel$.

The velocity of any arbitrary material point on the material line, $\boldsymbol{u}(\lambda)$, can then be expressed in the frame spanned by ($\boldsymbol{\hat{e}}_\parallel$, $\boldsymbol{\hat{e}}_\perp)$ in two different ways by taking either direct time derivative of Eq. (\ref{eqn_geometry}) or the Taylor expansion based on the velocity information (see Supplemental Material). Comparing these two expressions for $\boldsymbol{u}(\lambda)$ leads to evolution equations for $r^s$ and $r^b_\perp$, which then yields the evolution equation for curvature of the material line
\begin{equation}\label{eqn_curv_evolution}
\begin{split}
    \frac{d\kappa}{dt}=&\left(\boldsymbol{\hat{e}}_\parallel\cdot\boldsymbol{H}\cdot\boldsymbol{\hat{e}}_\parallel\right)\cdot\boldsymbol{\hat{e}}_\perp\\
    &+\left(\boldsymbol{\hat{e}}_\perp\cdot\boldsymbol{S}\cdot\boldsymbol{\hat{e}}_\perp-2\boldsymbol{\hat{e}}_\parallel\cdot\boldsymbol{S}\cdot\boldsymbol{\hat{e}}_\parallel\right)\kappa.
\end{split}
\end{equation}
Here $\boldsymbol{S}$ and $\boldsymbol{H}$ are the rate-of-strain tensor and the velocity Hessian tensor following the trajectories of fluid elements, respectively.

\begin{figure}[h]
    \centering
    \includegraphics[width=0.9\linewidth]{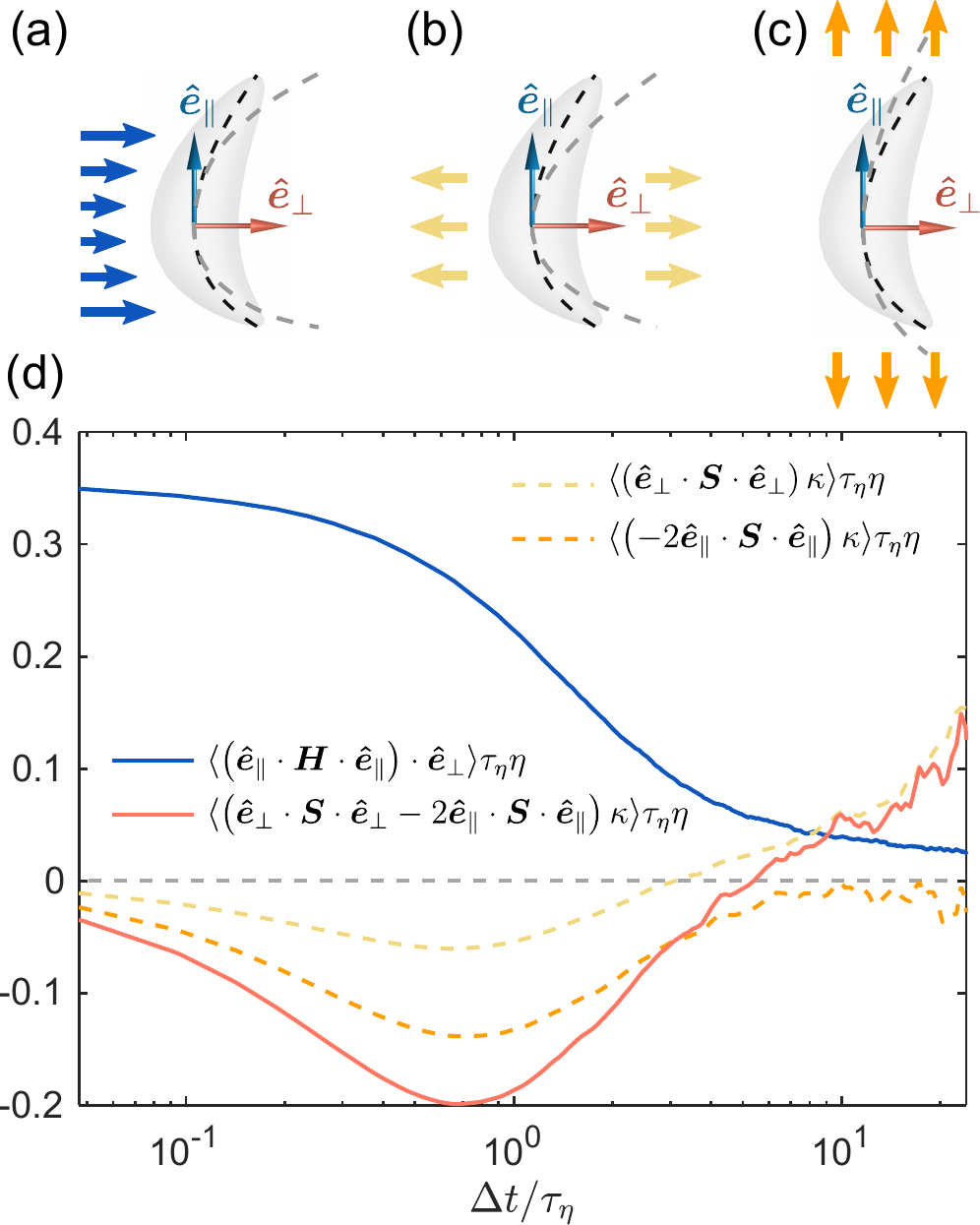}
    \caption{(a-c) Schematics illustrating how (a) velocity Hessian , (b) strain along $\boldsymbol{\hat{e}}_\perp$, and (c) strain along $\boldsymbol{\hat{e}}_\parallel$  contribute to the curvature change, respectively. For all cases, the black dashed curves represent the special material line (skeleton) while the gray dashed curves indicate the same material line at a later time deformed by the surrounding flows indicated by the thick arrows. (d) The time evolution of the contribution to the mean curvature growth by each term in Eq. (\ref{eqn_curv_evolution}), conditioned on $\kappa_1>3\langle\kappa_1\rangle$. All the terms are normalized by the Kolmogorov scales. }
    \label{fig_alignment}
\end{figure}

Eq. (\ref{eqn_curv_evolution}) holds for an arbitrary material line, so it also works for the curvature along the largest stretching ($\boldsymbol{\hat{e}}_{R1}$) direction $\kappa_1$. The first term on the right side of Eq. (\ref{eqn_curv_evolution}) represents the contribution from the velocity Hessian, which can directly bend the fluid element as shown in Fig. \ref{fig_alignment}(a). Here, the thick blue arrows indicate the primary velocity Hessian that bends the element (i.e., the velocity gradient that changes along the $\boldsymbol{\hat{e}}_\parallel$ direction). In the short time limit, $\kappa_1\rightarrow0$, all the terms multiplied by $\kappa_1$ in Eq. (\ref{eqn_curv_evolution}) are negligible, 
so Eq. (\ref{eqn_curv_evolution}) can be simplified to $d\kappa_1/dt=\left(\boldsymbol{\hat{e}}_\parallel\cdot\boldsymbol{H}\cdot\boldsymbol{\hat{e}}_\parallel\right)\cdot\boldsymbol{\hat{e}}_\perp$, which corresponds to the linear growth in the early stage as in Fig. \ref{fig_element_deformation}(b). At later times ($\Delta t>\tau_\eta$), this contribution of the velocity Hessian approaches zero as shown in Fig. \ref{fig_alignment}(d) (blue solid line) because  $\left(\boldsymbol{\hat{e}}_\parallel\cdot\boldsymbol{H}\cdot\boldsymbol{\hat{e}}_\parallel\right)$ may not be perfectly aligned with $\boldsymbol{\hat{e}}_\perp$. Since the velocity Hessian is a small-scale quantity, it is not surprising that the transition in Fig. \ref{fig_element_deformation}(b) begins at a small $\Delta t$ as the velocity Hessian decorrelates \cite{schumacher2007asymptotic}.

In addition to the Hessian term, the other two terms in Eq. (\ref{eqn_curv_evolution}), both proportional to $\kappa_1$, represent how the strain affects the curvature of an already-bent fluid element. Here, $\boldsymbol{\hat{e}}_\perp\cdot\boldsymbol{S}\cdot\boldsymbol{\hat{e}}_\perp$ represents the stretching along $\boldsymbol{\hat{e}}_\perp$, which tends to increase the curvature (as shown in Fig. \ref{fig_alignment}(b));  $\boldsymbol{\hat{e}}_\parallel\cdot\boldsymbol{S}\cdot\boldsymbol{\hat{e}}_\parallel$ represents the stretching along $\boldsymbol{\hat{e}}_\parallel$, which straightens an already-bent fluid element and reduces the curvature (as shown in Fig. \ref{fig_alignment}(c)). At later times, the mean curvature $\langle\kappa_1\rangle$ is large so both terms associated with $\kappa_1$ become dominant, leading to $d\kappa_1/dt\propto \kappa_1$. As a result, the late stage growth of curvature exhibits exponential trend, consistent with the results in Fig. \ref{fig_element_deformation}(b) inset.

The contributions from strain by each of the two terms (dashed line) and their combination (red solid line) are shown in Fig. \ref{fig_alignment}(d). The statistics were collected by only using the fluid elements with  $\kappa_1>3\langle\kappa_1\rangle$ because the late stage is dominated by the large-curvature cases as indicated by Eq. (\ref{eqn_curv_evolution}). It is evident that, as the velocity Hessian contribution approaches zero, the total contribution by the strain grows significantly, signaling the transition of the roles between these two mechanisms. This growing contribution by the strain is dominated by $(\boldsymbol{\hat{e}}_\perp\cdot\boldsymbol{S}\cdot\boldsymbol{\hat{e}}_\perp)\kappa$ which enhances the folding, whereas the other term  $(-\boldsymbol{\hat{e}}_\parallel\cdot\boldsymbol{S}\cdot\boldsymbol{\hat{e}}_\parallel)\kappa$ that reduces the curvature plateaus close to zero.

To understand the enhanced curvature intermittency at the late stage, the time evolution of the PDF of $\kappa_1$, i.e. $p(\kappa_1,t)$ as shown in Fig. \ref{fig_curv_pdf}(b), is modelled by assuming that $p(\kappa_1,t)d\kappa_1=p(\kappa_1',t+dt)d\kappa_1'$, where $\kappa_1'=\kappa_1+(d\kappa_1/dt)dt$ is the curvature of the fluid elements with an initial curvature $\kappa_1$ after $dt$. Substituting $\kappa_1'$ into the equation for PDF leads to,
\begin{equation}\label{eqn_pdf_evolution}
    \frac{\partial p}{\partial t}+(d\kappa_1/dt)\cdot\frac{\partial p}{\partial \kappa_1}+p\cdot\frac{d(d\kappa_1/dt)}{d\kappa_1}=0.
\end{equation}
Here we approximate $d\kappa_1/dt\approx\langle\boldsymbol{\hat{e}}_\perp\cdot\boldsymbol{S}\cdot\boldsymbol{\hat{e}}_\perp-2\boldsymbol{\hat{e}}_\parallel\cdot\boldsymbol{S}\cdot\boldsymbol{\hat{e}}_\parallel\rangle\kappa_1$ because (i) the strain is the dominant mechanism in the late stage and (ii) the contribution by velocity Hessian will only result in a self-similar distribution of curvature as shown in Fig. \ref{fig_curv_pdf}(a), whereas the PDFs in the late stage exhibit longer tails over time. Eq. (\ref{eqn_pdf_evolution}) is then solved numerically with  $p(\kappa_1)$ at $t/\tau_\eta=3$ obtained from the DNS data serving as the initial condition.

The predicted PDFs at different times are shown as the dashed curves in Fig. \ref{fig_curv_pdf}(b). An overall good agreement between the prediction and the data is achieved up to $t\approx10\tau_\eta$, particulary in the tail region extended beyond $\kappa_1\eta\approx 0.2$ in Fig. \ref{fig_curv_pdf}(b), which correspond to a length scale smaller than 5$\eta$. This suggests that the intermittency shown here is related to the curved elements being stretched even further by small-scale straining motions in the dissipative range. Note that the range of $\kappa_1\eta$ is limited because of the exceedingly low probability of finding fluid elements with $\kappa_1\eta$ greater than 0.25. We also note that the model following Eq. (\ref{eqn_pdf_evolution}) is simplified and it only holds when $d\kappa_1/dt$ increases with $\kappa_1$, i.e., more curved elements are being bent at a faster rate, which can only be satisfied at the late stage given the overall positive magnitude of $\langle\boldsymbol{\hat{e}}_\perp\cdot\boldsymbol{S}\cdot\boldsymbol{\hat{e}}_\perp-2\boldsymbol{\hat{e}}_\parallel\cdot\boldsymbol{S}\cdot\boldsymbol{\hat{e}}_\parallel\rangle$ in Eq. (\ref{eqn_curv_evolution}). Furthermore, the model is intended only for the tail region because the peak region with smaller $\kappa_1$ is dominated by the velocity Hessian. As a result, a mismatch between model predictions and simulation results is not unexpected for smaller $\kappa_1\eta$.



\begin{figure}
    \centering
    \includegraphics[width=0.9\linewidth]{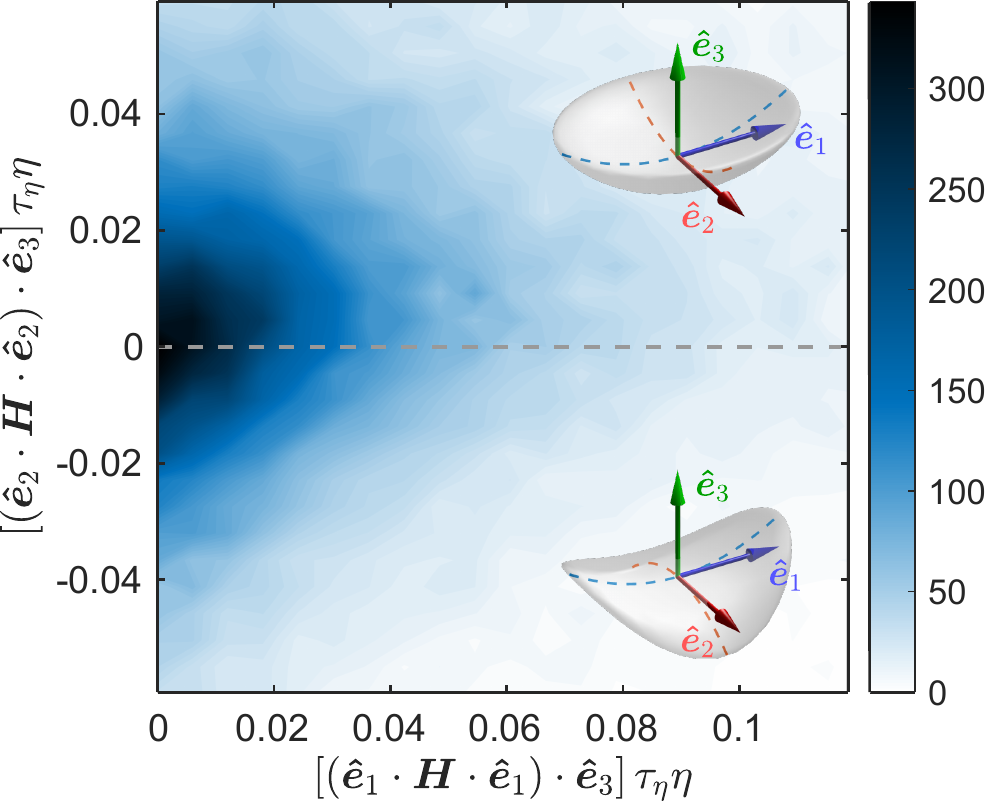}
    \caption{The joint PDF of the normalized curvature along $\boldsymbol{\hat{e}}_1$  and $\boldsymbol{\hat{e}}_2$ directions. Two schematics show an initially spherical fluid elements deforming to a bowl shape (top) and a saddle shape (bottom) after a short time,  respectively. }
    \label{fig_joint_pdf}
\end{figure}

Eq. (\ref{eqn_curv_evolution}) also enables us to use simple Eulerian quantities to understand folding in the early stage. As $\Delta t\rightarrow0$, $\boldsymbol{\hat{e}}_\parallel$ approaches $\boldsymbol{\hat{e}}_1$, which is the one of the three eigenvectors [$\boldsymbol{\hat{e}}_i$ ($i=1,2,3$)] corresponding to the maximum eigenvalue of the rate-of-strain tensor $\boldsymbol{S}$. The early growth of the material curvature can therefore be determined by an Eulerian quantity $\langle|\boldsymbol{\hat{e}}_1\cdot\boldsymbol{H}\cdot\boldsymbol{\hat{e}}_1|\rangle$ following $d\langle\kappa_1\rangle/dt\approx\langle\left(\boldsymbol{\hat{e}}_\parallel\cdot\boldsymbol{H}\cdot\boldsymbol{\hat{e}}_\parallel\right)\cdot\boldsymbol{\hat{e}}_\perp\rangle\approx \langle|\boldsymbol{\hat{e}}_1\cdot\boldsymbol{H}\cdot\boldsymbol{\hat{e}}_1|\rangle\beta$, where $\beta\approx0.85$ is the mean cosine of the angle between $\boldsymbol{\hat{e}}_\parallel\cdot\boldsymbol{H}\cdot\boldsymbol{\hat{e}}_\parallel$ and $\boldsymbol{\hat{e}}_\perp$ obtained from the DNS data.  The predicted result is shown as the cyan dashed line in Fig. \ref{fig_element_deformation}(b), and it overlaps with the DNS data perfectly. 



This Eulerian quantity $\langle|\boldsymbol{\hat{e}}_1\cdot\boldsymbol{H}\cdot\boldsymbol{\hat{e}}_1|\rangle$ also helps to establish a better physical picture of the deformed fluid elements in the short time limit beyond a simple flat sheet that extends along the $\boldsymbol{\hat{e}}_1$ and  $\boldsymbol{\hat{e}}_2$ directions considered in the classical framework \cite{lund1994improved}. As illustrated in the schematics of Fig. \ref{fig_joint_pdf}, such a sheet could be curved along $\boldsymbol{\hat{e}}_3$ direction, and its geometry can be described by two curvatures, whose growth are controlled by $(\boldsymbol{\hat{e}}_1\cdot\boldsymbol{H}\cdot\boldsymbol{\hat{e}}_1)\cdot\boldsymbol{\hat{e}}_3$ and $(\boldsymbol{\hat{e}}_2\cdot\boldsymbol{H}\cdot\boldsymbol{\hat{e}}_2)\cdot\boldsymbol{\hat{e}}_3$, respectively. 


The joint PDF of $(\boldsymbol{\hat{e}}_1\cdot\boldsymbol{H}\cdot\boldsymbol{\hat{e}}_1)\cdot\boldsymbol{\hat{e}}_3$ and $(\boldsymbol{\hat{e}}_2\cdot\boldsymbol{H}\cdot\boldsymbol{\hat{e}}_2)\cdot\boldsymbol{\hat{e}}_3$ normalized by Kolmogorov scales is shown in Fig. \ref{fig_joint_pdf}. Here, the direction of $\boldsymbol{\hat{e}}_3$ is chosen such that $(\boldsymbol{\hat{e}}_1\cdot\boldsymbol{H}\cdot\boldsymbol{\hat{e}}_1)\cdot\boldsymbol{\hat{e}}_3>0$, while $(\boldsymbol{\hat{e}}_2\cdot\boldsymbol{H}\cdot\boldsymbol{\hat{e}}_2)\cdot\boldsymbol{\hat{e}}_3$ can be either positive (bowl shape) or negative (saddle shape). The joint PDF suggests a nearly symmetric probability for either shape, skewing only slightly towards the bowl case.  Nevertheless, for a given curvature in one direction, the most likely curvature in the other direction is zero, so there appears to be some preference for cigar like shapes. This is confirmed in Fig. \ref{fig_element_deformation}(a) where the bending occurs mostly in one direction (although various other bending configurations can be seen).  Note that large values of the velocity Hessian may be the result of local instabilities (e.g. shear instabilities that are responsible for rolling up the vortex sheets into tubes \citep{vincent1994dynamics}). Connecting the dynamics
of instabilities to velocity Hessian and curvature requires further investigations.




In sum, our work establishes a new framework to connect folding dynamics to the velocity Hessian and deformation Hessian tensors in a way similar to the connection between stretching to velocity gradient  and Cauchy-Green strain tensors. As the stretching can be well described by the Lyapunov exponents based on strain, such a relationship may inspire the development of new ways to formulate the dynamical system for folding. Our framework also provides new insights into the flow intermittency that the  sharp-turning points in flows become even more curved due to strain, which could help gain deeper insights into the intermittency and inhomogeneity of turbulent mixing. Future work can possibly extend our framework to finite-sized fluid elements considering the coarse-graining effect at the same length scale. This extension will help develop improved models for length-scale reduction in the energy cascade process.


We acknowledge the financial support from the National Science Foundation under the award number CAREER-1905103. This project was also partially supported by the ONR award: N00014-21-1-2083.

\bibliography{ref}

\end{document}